\documentclass[runningheads]{llncs}
\usepackage{graphicx}
\usepackage{csquotes}
\usepackage{relsize}
\usepackage{adjustbox}
\usepackage{tikz}
\usepackage{pgfplots}
\usepackage{hyperref}
\usepackage{flushend} 
\usepackage{caption}
\usepackage{graphicx}
\usepackage{alltt}
\usepackage{csquotes}
\usepackage{float}
\floatstyle{ruled}
\newfloat{program}{t}{lop}
\floatname{program}{Program}
\newfloat{syntax}{t}{lop}
\floatname{syntax}{Syntax}
\usepackage{relsize}
\usepackage{adjustbox}
\usepackage{tikz}
\usepackage{pgfplots}
\usepackage{multicol}
\usepackage{multirow}
\usepackage{calc}
\usepackage{ifthen}
\usepackage{datapie}
\newcommand{\etal}{\textit{et al.}}
\newcommand{\smartinduct}{\texttt{smart\_induct}}

\newcommand{\induct}{the \texttt{induct} tactic}

\newcommand{\lifter}{\texttt{LiFtEr}}
\newcommand{\selfie}{\texttt{SeLFiE}}

\pgfplotsset{compat=1.17}
\usepackage{soul}
\usepackage[]{hyphenat}
\newcommand{\semanticinduct}{\texttt{sem\_ind}}

\usepackage{relsize}
\usepackage{adjustbox}
\usepackage{tikz}
\usepackage{pgfplots}
\usepackage{multicol}
\usepackage{calc}
\usepackage{ifthen}
\usepackage{graphicx}

\usepackage{booktabs}   
\usepackage{subcaption} 

\begin{document}
\title{Definitional Quantifiers Realise Semantic Reasoning for Proof by Induction}
\titlerunning{Definitional Quantifiers}
%
\author{Yutaka Nagashima\inst{1}\orcidID{0000-0001-6693-5325}}
%
\authorrunning{Yutaka Nagashima}
%
\institute{Independent researcher, Cambridge, UK\\
\email{united.reasoning@gmail.com}} 
%
\maketitle              
\begin{abstract}
Proof assistants offer tactics to apply proof by induction, 
but these tactics rely on inputs given by human engineers.
To automate this laborious process,
we developed \selfie{},
a boolean query language
to represent experienced users’ knowledge on how to apply 
\induct{} in Isabelle/HOL: 
when we apply an induction heuristic written in \selfie{} to an inductive problem and arguments to \induct{}, 
the \selfie{} interpreter judges 
whether the arguments are plausible for that problem according to the heuristic
by examining both the syntactic structure of the problem and definitions of the relevant constants.
To examine the intricate interaction between syntactic analysis and analysis of constant definitions,
we introduce \textit{definitional quantifiers}.
For evaluation
we build an automatic induction prover using \selfie{}.
Our evaluation based on 347 inductive problems 
shows that our new prover achieves $1.4 \cdot{} 10^{3}\%$ improvement over
the corresponding baseline prover for 1.0 second of timeout
and the median value of speedup is 4.48x.
\end{abstract}

\section{Introduction}\label{sec:intro}

The automation of proof by induction is a long standing challenge in Computer Science.
Conventionally, human researchers manually investigate both inductive problems and relevant definitions to decide how to apply proof by induction.
To mechanise such analysis, this paper introduces \textit{definitional quantifiers}:
quantifiers that range over the defining clauses of relevant constants to 
capture semantic properties of inductive problems.

\subsection{Motivating Example}\label{sec:motivating_example}

Consider the following two ways to define a reverse function for lists
presented in a tutorial of Isabelle/HOL \cite{isabelle}:

\begin{alltt}
@ :: \(\alpha\) list \(\Rightarrow\) \(\alpha\) list \(\Rightarrow\) \(\alpha\) list
  [] @ ys = ys 
| (x # xs) @ ys = x # (xs @ ys)

rev1 :: \(\alpha\) list \(\Rightarrow\) \(\alpha\) list 
  rev1 [] = [] 
| rev1 (x # xs) = rev1 xs @ [x]

rev2 :: \(\alpha\) list \(\Rightarrow\) \(\alpha\) list \(\Rightarrow\) \(\alpha\) list
  rev2 [] ys = ys 
| rev2 (x # xs) ys = rev2 xs (x # ys)
\end{alltt}

\noindent
where \verb|#| is the list constructor, 
\texttt{[x]} is a syntactic sugar for \texttt{x \# []},
and \verb|@| is the infix operator to append two lists into one.
How do you prove the following equivalence lemma?
\begin{alltt}
lemma "rev2 xs ys = rev1 xs @ ys"    
\end{alltt}
\noindent

Since both reverse functions are defined recursively,
it is natural to guess we can tackle this problem with proof by induction.
But how do you apply proof by induction to this inductive problem?
In this paper,
we present \selfie{}, 
a boolean query language
to encode
induction heuristics in a declarative form,
and its fast interpreter developed from scratch.
\selfie{} is embedded in Isabelle/ML, the implementation language of Isabelle/HOL,
and implemented for Isabelle2020.
The key idea behind \selfie{} is \textit{definitional quantifiers}:
new kinds of quantifiers that allow for definitional reasoning
in a domain-agnostic style.

\subsection{Background}
A prominent proof automation approach for proof assistants is the so-called hammer-style tools, such as 
HOL(y)
Hammer \cite{holy_hammer} for HOL-light \cite{hollight},
CoqHammer \cite{coqhammer} for Coq, and
Sledgehammer \cite{sledgehammer} for Isabelle/HOL \cite{isabelle}.
Sledgehammer, for example, translates proof goals in the polymorphic higher-order logic of Isabelle/HOL to monomorphic first-order logic and
attempts to prove the translated goals using various external automated provers.
Even though Sledgehammer brought powerful automation to Isabelle/HOL \cite{judgement_day};
when it comes to inductive theorem proving
the essence of inductive problems is lost in the translation,
severely impairing the performance of Sledgehammer.

This is unfortunate: 
most analyses of programs and programming languages involve reasoning about 
recursive data structures 
and procedures containing recursion or iteration \cite{alan1},
and inductive problems are essential to these analyses.



We address this long standing challenge with
\selfie{}. 
\selfie{} stands for
\underline{s}emantic-aware \underline{l}ogical \underline{f}eature \underline{e}xtraction.
\selfie{} has two main features:
\textit{definitional quantifiers}, 
and
\textit{domain-agnosticism}.
Domain-agnosticism allows users to encode induction heuristics that can transcend problem domains,
whereas definitional quantifiers allow \selfie{} heuristics to examine 
not only the syntactic structures of inductive problems but also the definitions of relevant constants.

Our implementation, available at GitHub \cite{GitHub},
is specific to Isabelle/HOL:
we implemented our system as an Isabelle theory for 
smooth user experience.
However, the underlying concept of definitional reasoning
is transferable to other proof assistants, such as 
Coq, Lean \cite{lean}, and HOL\cite{hol4}:
no matter which proof assistant we use,
we need to reason over
not only the syntactic structure of proof goals
but also definitions relevant to the goals
to decide how to apply proof by induction.
%

The rest of the paper is organized as follows.
Section \ref{sec:induction_in_isabelle} shows how to apply proof by induction
in Isabelle using the example from Section \ref{sec:motivating_example} and 
clarifies the need for reliable heuristics.
Section \ref{sec:overview} gives an overview of what we mean by encoding
induction heuristics and 
applying them to inductive problems in Isabelle/HOL.
Since it is still a new approach to reason over inductive problems
using a boolean query language,
Section \ref{sec:syntactic_reasoning} reviews \lifter{} \cite{lifter}, 
an existing framework developed 
to encode syntax-based heuristics for Isabelle/HOL. 
In particular, we observe how
\lifter{}'s quantifiers allow us to write heuristics
in a domain-agnostic style.
Then, we identify what induction heuristics we can\textit{not} encode in \lifter{}.
In Section \ref{sec:semantic_reasoning},
we present \selfie{} and its fast interpreter developed from scratch. 
In addition to the domain-agnosticism given by \lifter{}'s quantifiers, 
\selfie{} enables definitional reasoning using new language constructs that allow for
the reasoning about both the syntactic structure of proof goals and 
the definitions of relevant constants.
In Section, \ref{sec:evaluation} 
we introduce a recommendation system for \induct{} 
as a use case of \selfie{} and build a fast automatic inductive prover using this recommendation system,
and we discuss how much performance gain \selfie{} brought to 
inductive theorem proving in Isabelle/HOL.

\section{Proof by Induction in Isabelle/HOL}\label{sec:induction_in_isabelle}

Modern proof assistants come with \textit{tactic}s to facilitate proof by induction.
For example, Isabelle/HOL offers \induct{}.
The user-interface of \induct{} allows for 
an intuitive application of proof by induction.
For example, Nipkow \etal{} \cite{isabelle} proved our motivating example as follows:

\begin{verbatim}
lemma model_proof: "rev2 xs ys = rev1 xs @ ys"
 apply(induct xs arbitrary: ys) by auto
\end{verbatim}

That is to say, they firstly applied structural induction on \verb|xs| 
while generalizing \verb|ys|.
Since \verb|xs| is a list of any type, 
this application of structural induction resulted in the following two sub-goals:

\begin{alltt}
1. \(\forall{}\)ys. rev2 [] ys = rev1 [] @ ys
2. \(\forall{}\)a xs ys. (\(\forall{}\)ys. rev2 xs ys = rev1 xs @ ys) \(\Longrightarrow\)
  rev2 (a # xs) ys = rev1 (a # xs) @ ys
\end{alltt}
\noindent
where $\forall{}$ and $\Longrightarrow$ represent the universal quantifier and implication
of Isabelle's underlying logic respectively.
The first sub-goal is the base case for the structural induction,
whereas the second sub-goal is the step case where
we are asked to prove that 
this conjecture holds for \texttt{(a \# xs)} and \texttt{ys},
assuming that the conjecture holds for 
the same \texttt{xs} and an \textit{arbitrary} \texttt{ys}.
Then, they proved the remaining sub-goals using the general purpose tactic, \texttt{auto}.
For the step case, \texttt{auto} rewrote the left-hand side of the meta-conclusion as follows:
\begin{alltt}
   rev2 (a # xs) ys \hfill{} \(using the second clause defining\) rev2
\(\hookrightarrow\) rev2 xs (a # ys)
\end{alltt}
\noindent
whereas \texttt{auto} rewrote the right-hand side as follows:
\begin{alltt}
   rev1 (a # xs) @ ys \hfill{} \(using the second clause defining\) rev1
\(\hookrightarrow\) (rev1 xs @ [a]) @ ys \hfill{} \(using the associative property of\) @
\(\hookrightarrow\) rev1 xs @ ([a] @ ys) \hfill{} \(using the second clause defining\) @
\(\hookrightarrow\) rev1 xs @ (a # ([] @ ys)) \hfill{} \(using the first clause defining\) @
\(\hookrightarrow\) rev1 xs @ (a # ys)
\end{alltt}

\noindent
Applying such rewriting, \texttt{auto} internally transformed the step case to the following intermediate goal:
\begin{alltt}
\(\forall{}\)a xs ys. (\(\forall{}\)ys. rev2 xs ys = rev1 xs @ ys) \(\Longrightarrow\) 
  rev2 xs (a # ys) = rev1 xs @ (a # ys)
\end{alltt}

\noindent
Since \texttt{ys} was generalized in the induction hypothesis,
\texttt{auto} proved \texttt{rev2 xs (a} 
\verb|#| 
\texttt{ys) = rev1 xs} 
\verb|@| 
\texttt{(a}
\verb|#| 
\texttt{ys)}
by considering it as a concrete case of the induction hypothesis.
If Nipkow \etal{} had not passed \texttt{ys} to the \texttt{arbitrary} field,
\induct{} would have produced the following sub-goals:

\begin{alltt}
1. rev2 [] ys = rev1 [] @ ys
2. \(\forall{}\)a xs ys. (rev2 xs ys = rev1 xs @ ys) \(\Longrightarrow\)
  rev2 (a # xs) ys = rev1 (a # xs) @ ys
\end{alltt}

\noindent
This step case requests us to prove that
the original goal holds for \texttt{(a \# xs)} and \texttt{ys},
assuming that it holds for 
the same \texttt{xs} and the \textit{same} \texttt{ys} that appear in the induction hypothesis.
If we apply \texttt{auto} to these sub-goals,
\texttt{auto} proves the base case, but it leaves the step case as follows:
\begin{alltt}
\(\forall{}\)a xs. rev2 xs ys = rev1 xs @ ys \(\Longrightarrow\) 
  rev2 xs (a # ys) = rev1 xs @ (a # ys)
\end{alltt}

\noindent
That is, \texttt{auto} is unable to complete the proof attempt
because \texttt{ys} is shared both in the conclusion and induction hypothesis,
illustrating the importance of variable generalization.

Note that we did not have to develop induction principles manually
for \texttt{model\_proof}
since \induct{} found out how to apply structural induction 
from the arguments passed by Nipkow \etal{}
In fact, for most of the time Isabelle users do not have to develop induction principles manually,
but they only have to pass the right arguments to \induct{}.

Furthermore, 
there are often multiple equally appropriate ways to prove one theorem.
For example, we could have proved our running example with the following script:
\texttt{apply (induct xs ys rule: rev2.induct) by auto}.
This script applies computation induction using
the auxiliary lemma, \texttt{rev2.induct}, in the \texttt{rule} field.
Fortunately, in many cases
Isabelle automatically creates such auxiliary lemmas when defining relevant constants.
In our case,
Isabelle derived \texttt{rev2.induct} automatically when defining \texttt{rev2}.
This way, \induct{} reduces the problem of how to apply induction to
the following three questions:
 
\begin{itemize}
  \item On which terms do we apply induction?
  \item Which variables do we pass to the \texttt{arbitrary} field to generalize them?
  \item Which rule do we pass to the \texttt{rule} field?
\end{itemize}

However, answering these questions is a well-known challenge,
which used to require hard-won expertise.
We developed \selfie{} to encode such expertise.

\section{Overview of \selfie{}}\label{sec:overview}

\begin{figure}[t]
      \centerline{\includegraphics[width=0.8\linewidth]{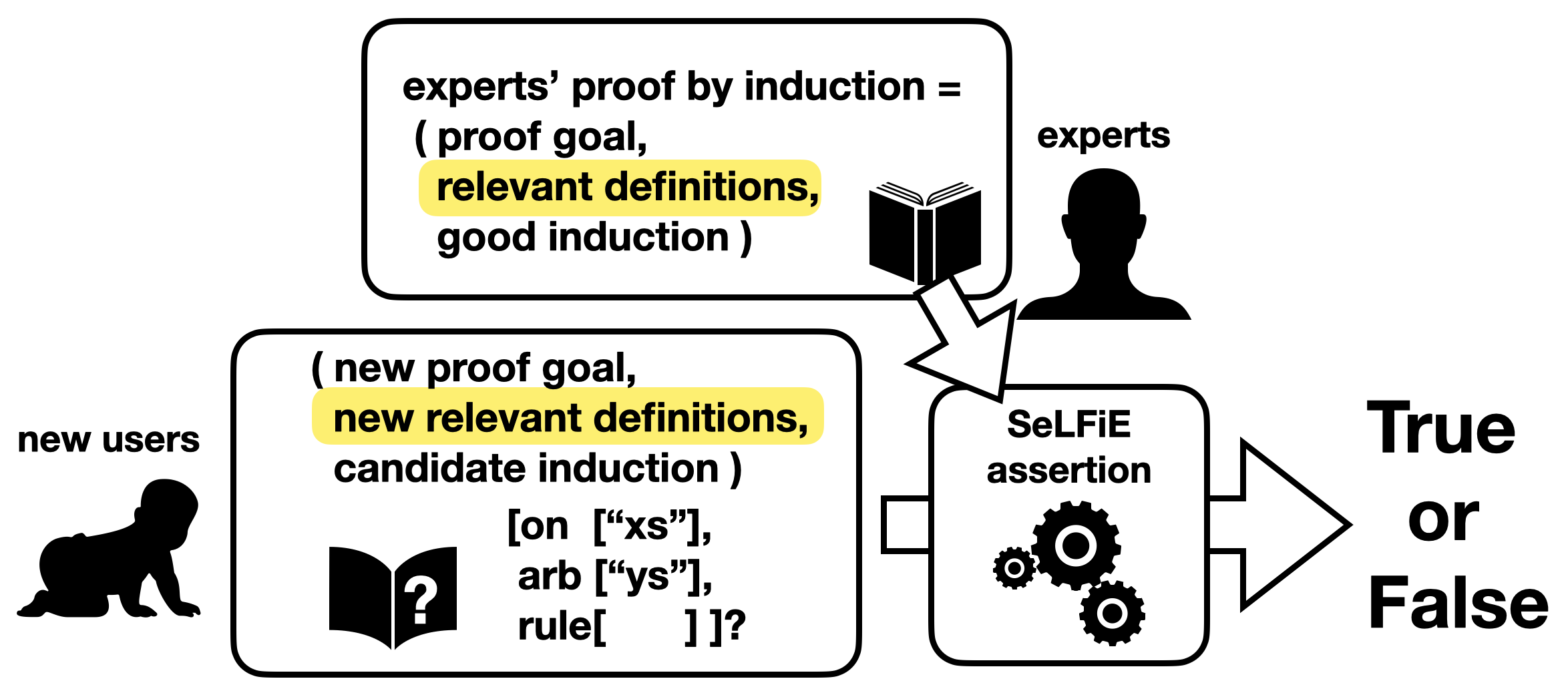}}
      \caption{The overview of \selfie{}.}
      \label{fig:overview}
\end{figure}

Figure \ref{fig:overview} shows how \selfie{} transfers such experienced users' knowledge
to new users:
when experienced users tackle inductive problems of their own,
they encode their expertise about how they use \induct{} as \selfie{} heuristics.
Each \selfie{} heuristic is an assertion that takes a triple of 
a proof goal, relevant constant definitions, and arguments passed to \induct{}.
A well-written \selfie{} assertion should return \texttt{True}
if the arguments to \induct{} are likely to be useful to prove the problem,
whereas it should return \texttt{False}
if the combination is not likely to be useful to prove the problem.
When new users want to know if their use of \induct{} is appropriate or not,
they apply the assertion written by an expert to their own problem
and learn if their choice of arguments is compatible with the induction heuristic
encoded by the expert.
Note that we \hl{highlighted} parts of Figure \ref{fig:overview} to emphasize
the main differences from the \selfie{}'s predecessor, \lifter{}, 
developed for a similar purpose.

Originally, we developed \selfie{}'s interpreter as an interactive tool to test a choice of proof by induction in terms of experts' heuristics. However, we can also use \selfie{} to build fully automated inductive provers as shown in Section \ref{sec:evaluation}.
In the following, we review \lifter{} and explain 
why we need a reasoning framework that can take relevant definitions
into account to encode reliable heuristics.

\section{Syntactic Reasoning in \lifter{}}\label{sec:syntactic_reasoning}

\subsection{\lifter{}: Logical Feature Extraction}\label{sec:lifter}

\begin{syntax}[!ht]
\begin{flushleft}
\textit{argument} \verb|:=| \textit{term} \texttt{|} \textit{number}\\
\textit{literal} \verb|:=| \textit{term\_occ} \texttt{|} \textit{rule} \texttt{|} \textit{argument} \texttt{|} \dots \\
\textit{assertion} \verb|:=| \textit{atomic} \texttt{|} \textit{literal} \texttt{|} \textit{connective} \texttt{|} \textit{quantifier}
\verb| |\texttt{|} \texttt{(} \textit{assertion} \verb|)| \\
\hspace{5.0em}\texttt{|} \hl{$\lambda$ \textit{assertions}. \textit{assertion}}
\verb| |\texttt{|} \hl{\textit{assertion assertions}}\\
\textit{type} \texttt{:=} \texttt{term} \texttt{|} \texttt{term\_occ} \texttt{|} \texttt{rule} \texttt{|} \texttt{number}\\
\textit{modifier} \texttt{:=}
\texttt{induction | arbitrary | rule}\\
\textit{quantifier} \verb|:=| $\exists$$x$ \verb|:| \textit{type}\verb|.| \textit{assertion}
\texttt{|} $\forall$$x$ \verb|:| \textit{type}\verb|.| \textit{assertion}\\
\hspace{5.2em}\texttt{|} $\exists x$ \verb|:| \textit{term} $\in$ \textit{modifier}\verb|.| \textit{assertion} 
\texttt{|} $\forall x$ \verb|:| \textit{term} $\in$  \textit{modifier}\verb|.| \textit{assertion}\\
\hspace{5.2em}\texttt{|} $\exists$$x$ \verb|:| \textit{term\_occ} $\in$ $y$ \verb|: term.| \textit{assertion} \\
\hspace{5.2em}\texttt{|} $\forall x$ \verb|:| \textit{term\_occ} $\in$ $y$ \verb|: term.| \textit{assertion}\\
\hspace{5.2em}\texttt{|} \hl{$\exists_D$\texttt{(} \textit{term} \texttt{,} 
    \textit{$\lambda$ \textit{arguments}. \textit{assertion}} \texttt{,} 
    \textit{arguments} \texttt{)}}\\
\hspace{5.2em}\texttt{|} \hl{$\forall_D$\texttt{(} 
    \textit{term} \texttt{,} 
    \textit{$\lambda$ \textit{arguments}. \textit{assertion}} \texttt{,} 
    \textit{arguments} \texttt{)}}\\
\textit{connective} \texttt{:= True | False}
\verb| |\texttt{|}\textit{assertion $\lor$ assertion} \texttt{|} \textit{assertion $\land$ assertion}
\verb| |\texttt{|}\textit{assertion $\rightarrow$ assertion} \texttt{|} $\neg$ \textit{assertion}\\
\textit{atomic} \verb|:=| \texttt{term\_is\_free (} \textit{term} \verb|)|\\
\hspace{3.8em}\texttt{|} \verb|are_same_term (| \textit{term} \verb|,| \textit{term} \verb|)|\\
\hspace{3.8em}\texttt{|} \verb|is_nth_argument_of|
\verb|(| \textit{term\_occ}\verb|, |\textit{number}\verb|, |\textit{term\_occ} \verb|)|\\
\hspace{3.8em}\texttt{|} \verb|is_nth_argument_in|
\verb|(| \textit{term\_occ}, \textit{number}, \textit{term\_occ} \verb|)|\\
\hspace{3.8em}\texttt{|} \verb|are_of_same_term (| \textit{term\_occ} \verb|,| \textit{term\_occ} \verb|)| \texttt{|} \texttt{\dots}\\
\end{flushleft}
\caption{The abstract syntax of \lifter{} / \selfie{} in one.
         The language 
         components unique to \selfie{} are \hl{highlighted}.
         }
\label{p:syntax}
\end{syntax}

\lifter{} is the first framework designed to describe how to use \induct{}
without relying on domain-specific constructs.
Syntax \ref{p:syntax} outlines \lifter{}'s syntax, 
which resembles that of first-order logic.
When reading Syntax \ref{p:syntax},
we ignore \hl{highlighted} parts,
which we discuss in Section \ref{sec:selfie}.

As shown in Syntax \ref{p:syntax}, 
\lifter{} offers four primitive variable types:
natural numbers, 
induction rules,
terms, and term occurrences.
An induction rule is an auxiliary lemma passed to the \texttt{rule} field of \induct{}.
The domain of terms is the set of all sub-terms appearing in the inductive problem at hand,
whereas the domain of term occurrences is the set of all occurrences of such sub-terms.
\lifter{} distinguishes terms and term occurrences explicitly
because we often have multiple distinct occurrences of the same term
in a syntax tree and have to analyze the locations of such occurrences.
For instance, the variable \texttt{ys} appears twice
in our theorem about list reversal.
But what matters when deciding which variables to generalize
is the occurrence of \texttt{ys} on the left-hand side
and its location relative to the only occurrence of \texttt{rev2},
as we shall see in Section \ref{sec:selfie_generalization_heuristic}.
Quantifiers over terms can be restricted to those terms that appear as arguments to
\induct{} under consideration.

\subsection{Naive Generalization Heuristic in \lifter{}}\label{sec:naive_generalization_heuristic_in_lifter}
As we saw in Section \ref{sec:induction_in_isabelle}, 
the key to the successful application of \induct{} for our motivating example is the generalization of 
\texttt{ys} using the \texttt{arbitrary} field.
When explaining why they decided to generalize \texttt{ys},
Nipkow \etal{} introduced the following generalization heuristic \cite{concrete_semantics}:

\begin{displayquote}
Generalize induction by generalizing all free variables (except the induction variable itself).
\end{displayquote}

\noindent
We can encode this generalization heuristic in \lifter{}
as shown in Program \ref{p:generalize_all_variables_that_are_not_induction_term}.
In plain English, Program \ref{p:generalize_all_variables_that_are_not_induction_term} reads as follows:

\begin{displayquote}
For any term, \textit{free\_var}, in a proof goal,
if \textit{free\_var} is a free variable 
but not passed to \induct{} as an induction term,
there exists a term, \textit{generalized},
in the \texttt{arbitrary} field
such that \textit{free\_var} and \textit{generalized} are the same term.
\end{displayquote}

\begin{program}[t]
\begin{alltt}
\(\forall\) \(free_var\) : term. 
   term_is_free (\(free_var\))
  \(\land\)
   \(\neg\) \(\exists\) \(induct\) : induction. are_same_terms (\(free_var\), \(induct\))
 \(\longrightarrow \)
  \(\exists\) \(generalized\) : arbitrary. are_same_terms (\(free_var\), \(generalized\))      
\end{alltt}
\caption{Naive generalization heuristic in \lifter{}}
\label{p:generalize_all_variables_that_are_not_induction_term}
\end{program}

\noindent
If we evaluate this heuristic for 
our ongoing example 
and its model proof by Nipkow \etal{},
the \lifter{} interpreter returns \texttt{True}, approving the generalization of \texttt{ys}.
But this heuristic seems too coarse to 
produce reliable recommendations.
In fact, Nipkow \etal{} articulate the limitation of this heuristic:

\begin{displayquote}
However, it (this generalization heuristic) should not be applied blindly. 
It is not always required, 
and the additional quantifiers can complicate matters in some cases. 
The variables that need to be quantified are typically those that change in recursive calls.
\end{displayquote}




Unfortunately, 
it is not possible to encode this provision in \lifter{}
because it involves reasoning on the structure of the syntax tree
representing the definition of 
a constant appearing in a proof goal,
which is \texttt{rev2} in this particular case.
In other words,
\lifter{} heuristics can describe the structures of proof goals
in a domain-independent style,
but they cannot describe the structures of relevant constants' definitions.
What is much needed is a framework to reason about
both arbitrary proof goals and their relevant definitions
in terms of the arguments passed to \induct{}
in a domain-agnostic style.
And this is the main challenge addressed by \selfie{}.





\section{Semantic Reasoning in \selfie{}}\label{sec:semantic_reasoning}
\subsection{Semantics-Aware Logical Feature Extraction}\label{sec:selfie}

We designed \selfie{} to overcome \lifter{}'s limitation 
while preserving its capability to transcend problem domains.
Syntax \ref{p:syntax} presents the abstract syntax of \selfie{}.
Since \selfie{} inherits design choices from \lifter{}, 
we re-use Syntax \ref{p:syntax};
however, 
we now 
include the \hl{highlighted} constructs into our consideration.

Compared to \lifter{}, which resembles first-order logic,
\selfie{} adopts lambda abstractions and function applications
to support the \textit{definitional quantifiers},
$\exists{}_D$ and $\forall{}_D$.
These new quantifiers range over definitions of constants,
so that we can handle constant definitions abstractly
to develop semantic-aware induction heuristics that can transcend problem domains,
whereas the conventional quantifiers from \lifter{} range over
terms and term occurrences,
so that we can handle terms and their occurrences abstractly
to develop syntax-based induction heuristics in a domain-agnostic style.

More specifically,
each definitional quantifier takes a triple of:
\begin{itemize}
    \item a term whose defining clauses are to be examined,
    \item a lambda function, which examines the relevant definitions, and
    \item a list of arguments, 
    each of which is either a term or natural number. 
    They are passed to the aforementioned lambda function to bridge the gap between the analysis of a proof goal and 
    the analysis of relevant definitions.
\end{itemize}

\noindent
For example,
$\exists_D$ \texttt{(const,} $\lambda$\texttt{xs. f xs, as)}
returns \texttt{True} if
$\lambda$\texttt{xs. f xs}
returns \texttt{True} when applied to \texttt{as}
for \textit{at least one} clause that defines \texttt{const}.
Similarly, $\forall_D$ \texttt{(const,} $\lambda$\texttt{xs. f xs, as)} returns \texttt{True} if
$\lambda$\texttt{xs. f xs}
returns \texttt{True} when applied to \texttt{as}
for \textit{all} clauses that define \texttt{const}.

The conventional quantifiers outside and inside definitional quantifiers behave differently:
inside the lambda function passed as the second argument to definitional quantifiers,
conventional quantifiers' domains are based on the relevant definitions under consideration.
For example, a quantifier over terms inside a definitional quantifier ranges over
terms that appear in the relevant defining clause under consideration.

\begin{figure}[t]
      \centerline{\includegraphics[width=0.8\linewidth]{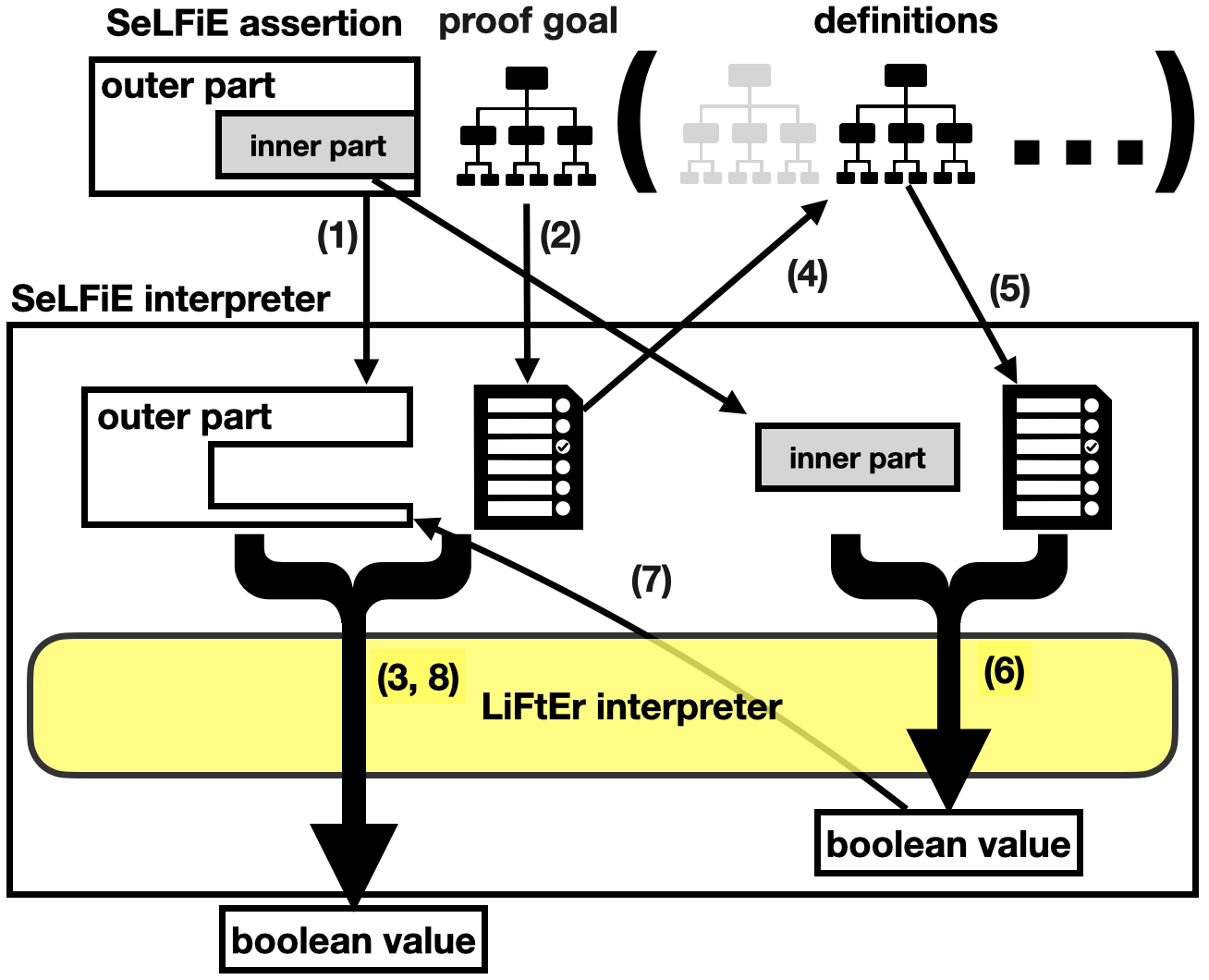}}
      \caption{The workflow of the \selfie{} interpreter.}
      \label{fig:selfie_interpreter}
\end{figure}

In the following
we focus on the operational aspect of definitional quantifiers, 
so that readers can grasp their nature
using a concrete example in Section \ref{sec:selfie_generalization_heuristic}.

Figure \ref{fig:selfie_interpreter} illustrates the overall workflow of the \selfie{} interpreter
when applied to an inductive problem and arguments of \induct{}.
In this figure, 
we assume that the \selfie{} assertion has only one definitional quantifier
for a simpler explanation;
however, in general, a \selfie{} heuristic may contain multiple definitional quantifiers. 
The small square, labelled as \texttt{inner part}, represents 
the lambda function passed as
the second argument to this definitional quantifier, whereas
\texttt{outer part} represents everything else
in the \selfie{} assertion.
Now based on this figure 
we explain how the \selfie{} interpreter works 
using the following eight steps from S1 to S8.

\begin{itemize}
    \item[S1.] Firstly, the \selfie{} interpreter takes a \selfie{} heuristic.
    \item[S2.] Then, the preprocessor of \selfie{} transforms the syntax tree representing the                                 inductive problem into a look-up table. 
               This look-up table replaces slow traversals in the syntax tree 
               with quick accesses to term occurrences 
               using their paths from the root node.
    \item[S3.] The \selfie{} interpreter processes the outer part of the assertion 
                  using the newly implemented \lifter{} interpreter. 
    \item[S4.] When the \selfie{} interpreter reaches the definitional quantifier,
                  it extracts the clauses that define the first argument 
                  of the definitional quantifier from the underlying proof context.
    \item[S5.] The interpreter transforms the syntax tree representing the relevant definitions into
                  look-up tables.
    \item[S6.] The \lifter{} interpreter applies the inner part of the assertion,
                  which is the lambda function passed as the second argument of the definitional quantifier,
                  to the list of arguments, which is the third argument of the definitional quantifier,
                  based on the look-up tables produced in S5.
    \item[S7.] The result of S6 is then returned to the \lifter{} interpreter.
    \item[S8.] The \lifter{} interpreter continues to evaluate the remaining outer part
               using the return value from the inner part.
\end{itemize}

We named our language \selfie{}
partly because 
we extended \lifter{}, 
so that \lifter{} can call it\underline{self}
to support definitional quantifiers,
but also because
\selfie{} heuristics can attain the \underline{s}emantics of 
inductive problems 
using definitional quantifiers.
Our motto is that:
\begin{displayquote}
We analyze inductive problems \textit{semantically} by 
analyzing their relevant definitions \textit{syntactically}.
\end{displayquote}
\noindent

\subsection{Semantics-Aware Generalization Heuristic}\label{sec:selfie_generalization_heuristic}


\begin{program}
\begin{alltt}
 \(\forall\) \(arb_term\) : term \(\in\) arbitrary.
  \(\exists\) \(f_term\) : term.
   \(\exists\) \(f_occ\) : term_occ \(\in\) \(f_term\).
    \(\exists\) \(arb_occ\) \(\in\) \(arb_term\).
     \(\exists\) \(generalize_nth\) : number.
       is_nth_argument_of (\(arb_occ\), \(generalize_nth\), \(f_occ\))
      \(\land\)
       \(\exists\sb{D}\)(\(f_term\), generalize_nth_argument_of, [\(generalize_nth\), \(f_term\)])
\end{alltt}
\caption{Syntactic analysis of more reliable generalization heuristic in \selfie{}}
\label{p:generalization_heuristic_in_selfie_outer}
\end{program}

\begin{program}
\begin{alltt}
generalize_nth_argument_of :=
 \(\lambda\) [\(generalize_nth\), \(f_term\)].
  \(\exists\) \(lhs_occ\) : term_occ. is_left_hand_side (\(lhs_occ\))
  \(\land\)
   \(\exists\) \(nth_param_on_lhs\) : term_occ.
     is_nth_argument_in (\(nth_param_on_lhs\), \(generalize_nth\), \(lhs_occ\))
    \(\land\)
     \(\exists\) \(nth_param_on_rhs\) : term_occ.
       \(\neg\) are_of_same_term (\(nth_param_on_rhs\), \(nth_param_on_lhs\))
      \(\land\)
       \(\exists\) \(f_occ_on_rhs\) : term_occ \(\in\) \(f_term\).
        is_nth_argument_of (\(nth_param_on_rhs\), \(generalize_nth\), \(f_occ_on_rhs\))
\end{alltt}
\caption{Definitional analysis of a generalization heuristic in \selfie{}}
\label{p:generalization_heuristic_in_selfie_inner}
\end{program}

We now improve the naive generalization heuristic from Section \ref{sec:naive_generalization_heuristic_in_lifter}
in \selfie{}.
More specifically,
we encode the provision to the generalization heuristic discussed in Section \ref{sec:naive_generalization_heuristic_in_lifter} as
Program \ref{p:generalization_heuristic_in_selfie_outer} and
Program \ref{p:generalization_heuristic_in_selfie_inner}.
Intuitively, when applied to \texttt{model\_proof}, 
these programs realise the following dialogue:
\begin{itemize}
    \item Program \ref{p:generalization_heuristic_in_selfie_outer} asks 
    ``Should we generalize \textit{ys}, which appears as the second argument of \texttt{rev2}?''
    \item Program \ref{p:generalization_heuristic_in_selfie_inner} answers 
    ``Yes, because the second argument changes from the left-hand side to the right-hand side in the second clause defining \texttt{rev2}.''
\end{itemize}
Keeping this dialogue in mind,
we examine how the \selfie{} interpreter formally processes this heuristic
for our running example.

\begin{itemize}
    \item [S1.] We pass 
    Program \ref{p:generalization_heuristic_in_selfie_outer} and  \ref{p:generalization_heuristic_in_selfie_inner},
    and \texttt{model\_proof} to the \selfie{} interpreter.
    \item [S2.] The interpreter transforms the syntax tree representing the proof goal into a look-up table
    for faster processing.
    \item [S3.] 
    The \selfie{} interpreter processes the outer part
    for the syntax tree representing the proof goal itself.
    Note that the domains of quantifiers over terms and term occurrences are based on
    those terms and their occurrences within 
    the proof goal itself.
    
    In \texttt{model\_proof}, only one variable, \texttt{ys}, is
    generalized in the \texttt{arbitrary} field. 
    Therefore, for \texttt{model\_proof} to satisfy this generalization heuristic
    we only have to satisfy inner existential quantifiers when 
    $arb\_term$ is \texttt{ys}.
    Thus, we instantiate each existentially quantified variable in Program \ref{p:generalization_heuristic_in_selfie_outer} as follows:
    \begin{itemize}
        \item \textit{f\_term} with \texttt{rev2},
        \item \textit{f\_occ} with the sole occurrence of \texttt{rev2} in the proof goal,
        \item \textit{arb\_occ} with the occurrence of \texttt{ys} on the left-hand side in the goal, and
        \item \textit{generalize\_nth} with $2$.
    \end{itemize}
    Then, \texttt{is\_nth\_argument\_of} returns \texttt{True}
    since \texttt{ys} on the left-hand side is the second argument to \texttt{rev2} in the goal.
    \item [S4.] When the interpreter hits $\exists_D$ with 
    \textit{f\_term} being \texttt{rev2},
    it extracts the two syntax trees defining \texttt{rev2} from the proof context.
    Since $\exists_D$ is an existential quantifier,
    we only have to show that
    Program \ref{p:generalization_heuristic_in_selfie_inner} returns \texttt{True}
    for one of the two equations defining \texttt{rev2}.
    In the following, we focus on the second clause, \texttt{rev2 (x \# xs) ys = rev2 xs (x \# ys)}.
    \item [S5.] The interpreter transforms 
    each syntax tree representing a clause 
    defining \texttt{rev2} into a look-up table for faster processing.
    \item [S6.] The interpreter evaluates Program \ref{p:generalization_heuristic_in_selfie_inner}
    with $2$ as \textit{generalize\_nth} and 
    \texttt{rev2} as \textit{f\_term},
    since they are passed from Program \ref{p:generalization_heuristic_in_selfie_outer}.
    Note that the domains of quantifiers over terms and term occurrences
    are now all terms and term occurrences in \texttt{rev2 (x \# xs) ys = rev2 xs (x \# ys)}.
    To satisfy Program \ref{p:generalization_heuristic_in_selfie_inner}
    we instantiate existentially quantified variables as follows:
    \begin{itemize}
        \item \textit{lhs\_occ} with the left-hand side of the equation, \texttt{rev2 (x \# xs) ys},
        \item \textit{nth\_param\_on\_lhs} with the occurrence of \texttt{ys},
        which appears as the second argument on the left-hand side,
        \item \textit{f\_occ\_on\_rhs} with the sole occurrence of \texttt{rev2} on the right-hand side, and
        \item \textit{nth\_param\_on\_rhs} with the sole occurrence of \texttt{x \# ys}, which is the second argument to \texttt{rev2} bound by \textit{f\_occ\_on\_rhs}.
    \end{itemize}
    Since \texttt{x \# ys} and \texttt{ys} are not the same term,
    the interpreter evaluates Program \ref{p:generalization_heuristic_in_selfie_inner}
    to \texttt{True} for the second clause defining \texttt{rev2},
    which is tantamount to say
    \textit{we generalize the second argument of \texttt{rev2} because the second argument of \texttt{rev2} changes in a recursive call}
    in a domain-agnostic style.
    \item [S7.] 
    Program \ref{p:generalization_heuristic_in_selfie_inner} returns
    \texttt{True} to Program \ref{p:generalization_heuristic_in_selfie_outer}.
    \item [S8.] With this returned value,  
    the interpreter evaluates Program \ref{p:generalization_heuristic_in_selfie_outer} to \texttt{True}.
\end{itemize}

This is how Program \ref{p:generalization_heuristic_in_selfie_inner}
encodes the provision to the generalization heuristic discussed in Section \ref{sec:naive_generalization_heuristic_in_lifter}.
Note that
the interaction between the two programs involves
natural numbers, terms, and boolean values only:
more complex reasoning, such as quantification over natural numbers, terms, and term occurrences,
happens only within each program
because each module has its own domains for terms and term occurrences.
Furthermore, it is not allowed to pass term occurrences 
from a syntactic analysis to a definitional analysis through definitional quantifiers.
Therefore, we discuss relative locations of certain term occurrences across syntax trees,
by passing natural numbers and terms from the syntax level
to the definition level, as is done in this example.
This clear separation between syntactic and definitional reasoning
improves the readability of this heuristic.

In this particular example, 
we demonstrated two-level analysis of syntax trees using two \selfie{} programs.
However, \selfie{}'s definitional quantifiers can orchestrate 
reasoning on arbitrary number of levels.

\section{Case Studies and Evaluations}\label{sec:evaluation}

\subsection{Interactive Recommendation System}

Using \selfie{}, we previously developed \semanticinduct{}, 
an interactive recommendation system 
for proof by induction in Isabelle/HOL \cite{faster_smarter}.
Given an inductive problem,
\semanticinduct{} produces a number of induction candidates
and applies 44 \selfie{} heuristics to these candidates.
Each heuristic is tagged with
a certain point, representing the weight of each heuristic.
Based on the sum of these points,
\semanticinduct{} ranks the candidates and presents the 10 most promising ones
to its users.

Nagashima evaluated \semanticinduct{} against 1,095 inductive proofs 
from the Archive of Formal Proofs (AFP) \cite{AFP} and 
compared \semanticinduct{} against its predecessor, \smartinduct{} \cite{smart_induction},
which is written in \lifter{}.

Table \ref{table:semantic_induct_coincidence} summarizes
how often \semanticinduct{}'s recommendations coincide with the choices of human engineers.
For example, 
Table \ref{table:semantic_induct_coincidence} shows 38.2\% for
``\semanticinduct{}'' at ``top 1''.
This means when considering only the top one candidate recommended by \semanticinduct{},
\semanticinduct{}'s recommendations coincide with the choices of human engineers 
for 38.2\% of proof goals in the dataset.
This is a 90.0\% improvement compared to \smartinduct{}, 
which reported 20.1 \% for ``top 1''.

Table \ref{table:semantic_induct_return}, on the other hand, summarizes
how long it takes for \semanticinduct{} to produce recommendations.
For example, 
Table \ref{table:semantic_induct_return} shows 8.8\% for
``\semanticinduct{}'' at ``0.2''.
This means 
\semanticinduct{} managed to produce recommendations for 8.8\% of proof goals in the dataset within 0.2 seconds of timeout.
Furthermore, Nagashima also reported that
the median value of the execution time of \semanticinduct{} is 
1.06 seconds, while
that of \smartinduct{} is 2.79 seconds,
which is a 2.63x speedup.

\begin{table}[t]
\begin{subtable}[t]{0.5\textwidth}
\begin{tabular}{c | r r r r}
\hline
\noalign{\smallskip}
tool & top 1 & top 3 & top 5 & top 10\\
\hline
\semanticinduct{} & 38.2 & 59.3 & 64.5 & 72.7\\ 
\smartinduct{}    & 20.1 & 42.8 & 48.5 & 55.3\\ 
\hline
\end{tabular}
\caption{Coincidence rates [\%]
}\label{table:semantic_induct_coincidence}
\end{subtable}
\begin{subtable}[t]{0.5\textwidth}
\begin{tabular}{c | r r r r r}
\hline
\noalign{\smallskip}
tool & 0.2& 0.5 & 1.0 & 2.0 & 5.0\\
\hline
\semanticinduct{} & 8.8 & 24.7 & 47.8 & 69.8 & 86.8 \\ 
\smartinduct{}    & 0.0 & 3.5  & 16.9 & 38.3 & 70.2 \\ 
\hline
\end{tabular}
\caption{Return rates [\%] within timeouts [s] 
}\label{table:semantic_induct_return}
\end{subtable}
\caption{Coincidence rates and return rates}
\end{table}

\subsection{Automatic Proof Search using \selfie{}}

\begin{program}[t]
\begin{alltt}
Auto_Solve = Thens[Auto, Solved]
PSL_WO_SeLFiE =
Ors[Auto_Solve,
    PThenOne[Dynamic (Induct), Auto_Solve]
    PThenOne[Dynamic(Induct), Thens[Auto, RepeatN(Hammer), Solved]]]
\end{alltt}
\caption{Automatic inductive prover without \selfie{}}
\label{p:psl_wo_selfie}
\end{program}

\begin{program}[t]
\begin{alltt}
PSL_W_SeLFiE =
Ors[Auto_Solve,
    PThenOne[\hl{Semantic\_Induct}, Auto_Solve]
    PThenOne[\hl{Semantic\_Induct}, Thens[Auto, RepeatN(Hammer), Solved]]]
\end{alltt}
\caption{Automatic inductive prover with \selfie{}}
\label{p:psl_w_selfie}
\end{program}

We integrated \texttt{sem\_ind} into an automatic inductive prover written in PSL \cite{psl}
and measured how \selfie{} improved PSL's automatic proof search.
PSL is a domain-specific language to describe rough ideas about 
how to find a proof using backtracking search 
over tactics in Isabelle/HOL.
In the following, we focus on PSL's constructs
used in our evaluation leaving out irrelevant details of PSL.

Program \ref{p:psl_wo_selfie} shows an example automatic inductive prover written in PSL, which we use as the baseline prover in this evaluation.
The strategy is called \texttt{PSL\_WO\_SeLFiE},
and it combines three sub-strategies using the deterministic combinator \texttt{Ors}:
it first tries the first sub-strategy, \texttt{Auto\_Solve}, 
and proceeds to the second sub-strategy only if the first sub-strategy fails, and so on.
\texttt{Thens} used in \texttt{Auto\_Solve} is the sequential combinator,
which combines \texttt{Auto} and \texttt{Solved} sequentially,
and \texttt{Auto} in PSL corresponds to the \texttt{auto} tactic in Isabelle, while
the following \texttt{Solved} checks if all sub-goals are proved by \texttt{auto}.
\texttt{Hammer} represents the invocation of Sledgehammer, which is
wrapped in \texttt{RepeatN} in Program \ref{p:psl_wo_selfie}.
This means 
``repeat applying Sledgehammer to the remaining sub-goals $n$ times where 
$n$ is the number of sub-goals before applying Sledgehammer''.
\texttt{PThenOne} is the sequential parallel combinator:
\texttt{PThenOne} takes exactly
two sub-strategies and applies the second sub-strategy to 
the results of the first sub-strategy in parallel
until at least one of them succeeds.

\texttt{Dynamic (Induct)}
creates variants of \induct{}s with different arguments based on the given goal
and combine such variants non-deterministically.
However, when the interpreter produces such variants of \induct{}s using \texttt{Dynamic (Induct)},
it does not know which one would be the most suitable induction.
Therefore, the interpreter naively combines variables and arguments
appearing in the proof goal to produce candidate \texttt{induct} tactics.
In PSL, it is the subsequent sub-strategies that are to identify 
the right arguments for \induct{}:
\texttt{PThenOne [Dynamic (Induct), Auto\_Solve]}, for example,
keeps applying \texttt{auto} to sub-goals emerging after applying
the \texttt{induct} tactic with various sequences of arguments 
until it finds a sequence that results in
sub-goals that are all proved by \texttt{auto}.


The drawback of this approach is that
PSL's interpreter cannot identify the appropriate arguments for the \texttt{induct}
tactic if it cannot complete a proof search:
for difficult inductive problems,
the interpreter often fails to complete a proof search within a realistic timeout
because \texttt{Dynamic (Induct)} tends to produce a large number of
induction candidates
and the necessary proof steps 
after applying \induct{} tend to be complicated.
What was lacking was the mechanism to identify promising induction candidates
without relying on a proof search,
so that PSL's interpreter can spend limited computational resources for
a small number of promising candidates to complete a proof search.
For this reason,
we integrated \semanticinduct{} into PSL,
and we counted how many goals are proved within each timeout.

Program \ref{p:psl_w_selfie} shows the new automatic prover.
Here, \texttt{Semantic\_Induct} represents \semanticinduct{} 
integrated into PSL's environment.
We highlighted the differences in Program \ref{p:psl_w_selfie} 
from Program \ref{p:psl_wo_selfie} to clarify that
we are using almost the same PSL strategy for a fair comparison
except for the introduction of \texttt{Semantic\_Induct}.

For our evaluation,
we used 12 Isabelle theory files from 8 projects about various topics 
in the AFP, which in total include 347 proofs by induction.
These projects are about 
the depth-first search \cite{dfs},
binomial heaps \cite{binomial},
a boolean expression checker \cite{boolean},
multi-dimensional binary search trees \cite{kd_tree},
the priority search tree \cite{priority_search_tree},
linear temporal logic \cite{LTL-AFP}, 
imperative programming language Simpl \cite{Simpl-AFP},
and program verification competition \cite{VerifyThis}.
We conducted this evaluation on a MacBook Pro (15-inch, 2019)
with 2.6 GHz Intel Core i7 6-core memory 32 GB 2400 MHz DDR4,
and the reported execution times are based on elapsed real time.

\begin{table}[t]
\begin{subtable}[t]{0.48\textwidth}
\centering
\begin{tabular}{c | c c}
\hline
\noalign{\smallskip}
timeouts & Program \ref{p:psl_w_selfie} & Program \ref{p:psl_wo_selfie}\\
\hline
0.3[s]  & 11.0\% & 1.2\%  \\ 
1.0[s]  & 25.6\% & 1.7\%  \\
3.0[s]  & 28.2\% & 21.9\% \\
10.0[s] & 34.9\% & 28.0\% \\
30.0[s] & 45.8\% & 38.3\% \\
\hline
\end{tabular}
\caption{Success rates 
}\label{table:psl_w_selfie}
\end{subtable}
\begin{subtable}[t]{0.48\textwidth}
\centering
\begin{tabular}{c | c}
\hline
\noalign{\smallskip}
speedup [times] &  occurrence\\
\hline
               x $<$ 1.0   & 3 (2.4\%) \\ 
1.0     $\leq$ x $<$ 5.0   & 64 (50.8\%) \\
5.0     $\leq$ x $<$ 10.0  & 44 (34.9\%) \\
10.0    $\leq$ x $<$ 15.0  & 9 (7.1\%) \\
15.0    $\leq$ x $<$       & 6 (4.8\%) \\
\hline
\end{tabular}
\caption{Speedup of execution time}\label{table:speedup_by_selfie}
\end{subtable}
\caption{Success rates and speedup}
\end{table}

Table \ref{table:psl_w_selfie} shows
how many inductive problems were proved by each program within each timeout.
For example,
the timeout of 0.3[s] for Program \ref{p:psl_w_selfie} has 11.0\%.
This means Program \ref{p:psl_w_selfie} proved 11.0\% inductive problems in the dataset within 0.3 seconds.
For a fair comparison
we included not only the time spent by tactics for proof search
but also 
the time spent by \semanticinduct{} 
when measuring the execution time of each proof search.
As shown in Table \ref{table:psl_w_selfie},
PSL enhanced with \semanticinduct{} proved more inductive problems than 
PSL without \semanticinduct{} for various timeouts.
For 30.0 seconds of timeout,
PSL with \semanticinduct{} proved 159 inductive problems,
while PSL without \semanticinduct{} proved 133 problems only.
126 problems were proved by both provers within this timeout.
For each problem proved by both programs within 30.0 seconds, 
we computed the speedup of execution time spent to complete each proof search.
For example, Program \ref{p:psl_w_selfie} spent 0.325 seconds 
and Program \ref{p:psl_wo_selfie} spent 2.171 seconds
to prove a lemma named \texttt{nexts\_set} in \texttt{DFS.thy}.
Therefore, the speedup of execution time for this lemma is
(2.171 / 0.325) = 6.68.

Table \ref{table:speedup_by_selfie} shows the distribution
of speedup observed among such problems.
For example, the second row reads 1.0 $\leq{}$ x $<$ 5.0 and 64.0 (50.8\%),
and this means that
Program \ref{p:psl_w_selfie} achieved between 1.0x to 5.0x speedup
compared to Program \ref{p:psl_wo_selfie}
for 64 inductive problems proved by both provers.
As shown in this table,
we confirmed that Program \ref{p:psl_w_selfie} achieved speedups over Program \ref{p:psl_wo_selfie} except for 3 cases,
which constitutes 2.4\% of problems proved by both
provers within 30.0 seconds of timeout.
The median value for speedup is 4.48x.

\section{Conclusion}

We presented \selfie{}, a boolean-query language to encode induction heuristics.
The abstraction brought by definitional quantifiers 
allow \selfie{} to transcend problem domains while analysing
not only the syntactic structures of
inductive problems but also definitions of relevant constants
in a modular style.


Our conservative extension to \lifter{}'s syntax allows us 
to take advantage of \lifter{}'s domain-agnosticism,
while adding the capability to reason on the semantics of proof goals.
To realise such extension, we implemented \selfie{}'s interpreter
from scratch: 
since \lifter{}'s original interpreter was not designed with 
definitional reasoning in mind, 
it did not support even lambda abstraction or function application,
and suffered from poor performance,
incremental improvement was not realistic.

Nagashima implemented \semanticinduct{} in \selfie{},
and we integrated \semanticinduct{} into PSL 
and built an automatic inductive prover.
Our experiment showed that
compared to the baseline prover
our inductive prover based on \selfie{} achieves
$1.4 \cdot{} 10^{3}\%$ improvement of success rate 
for 1.0 second of timeout
as well as a 4.48x speedup as the median value.

The final goal of this project is to build a strong inductive prover.
It remains our future work to further strengthen the automatic prover introduced in Section \ref{sec:evaluation},
by incorporating two conjecturing mechanisms, top-down conjecturing \cite{pgt} and bottom-up conjecturing \cite{hipster},
into our system.

\section{Related Work}

A well-known approach for inductive theorem proving is
the Boyer-Moore waterfall model \cite{waterfall},
which was invented for a first-order logic on Common Lisp \cite{lisp}.
In the original waterfall model,
a prover tries to apply any of the six techniques, including
simplification, generalization and induction.
If any of these techniques works, 
the prover stores the resulting sub-goals in a pool 
and continues to apply the techniques until it empties the pool.

ACL2 \cite{acl2} is the latest incarnation of this line of work
with industrial applications \cite{acl2_industry}.
To decide how to apply induction,
ACL2 estimates how good each induction scheme is
by computing a score, called \textit{hitting ratio},
based on a fixed formula \cite{acl_book,induction},
and it proceeds with the induction scheme with the highest hitting ratio.
Heras \etal{} used ML4PG learning method to find patterns
to generalize and transfer inductive proofs from one domain to another in ACL2
\cite{ml4pg_acl}.
Instead of computing a hitting ratio,
we provide \selfie{} as a language, so that
Isabelle experts can encode their expertise as assertions.

There are ongoing attempts to extend saturation-based superposition provers with induction:
Cruanes presented an extension of typed superposition
that can perform structural induction \cite{zipperposition},
while Reger \etal{} incorporated lightweight automated induction \cite{vampire_induction}
to the Vampire prover \cite{vampire}
and Hajdú \etal{} extended it to cover induction with generalization \cite{vampire_generalization}.
Contrary to their work, our approach to proof by induction uses Isabelle's default \verb|induct| tactic,
which we can use for arbitrary data types.

For more expressive logics,
Jiang \etal{} employed multiple waterfalls \cite{jiang} in HOL Light \cite{hollight}.
However, to decide induction variables, 
they naively picked the first free variable with recursive type and 
left the selection of promising induction variables as future work.
Passmore \etal{} developed the Imandra automated reasoning system \cite{imandra},
which also uses the waterfall model for its typed higher-order setting.
For Isabelle/HOL,
Dixon \etal{} developed IsaPlanner \cite{isaplanner},
a generic framework to encode proof plans \cite{proof_plan}.
IsaPlanner can incorporate reasoning techniques, such as rippling \cite{rippling},
for proof by induction.
For generalization, however,
IsaPlanner naively generalizes all non-induction variables \cite{isaplanner_phd}.

Machine learning tools for tactic-based theorem proving
mainly focus on tactic recommendations and premise selections,
leaving
the problem of arguments selection for tactics 
as an open question
when arguments are terms \cite{tactictoe,pamper,tactictoe_for_coq}. 
Instead of relying on machine learning algorithms,
we developed a language, in which one can explicitly encode heuristics.
We plan to use \selfie{} as a feature extractor for machine learning algorithms:
by applying \selfie{} heuristics to inductive problems,
we can convert each pair of an inductive problem and induction arguments to an array of boolean values,
which is amenable for machine learning algorithms.
The application of \selfie{} as a preprocessor for machine learning algorithms remains as our future work.

\section*{Acknowledgement}
We thank the anonymous reviewers for the useful feedback, both at Tests and Proofs 2022 and other conferences.
This work was supported by the following grants:
\begin{itemize}
    \item NII under NII-Internship Program 2019-2nd call,
    \item the European Regional Development Fund under the project AI $\&$ Reasoning. (reg.no.CZ.02.1.01/0.0/0.0/15\_003/0000466)
\end{itemize}

%
%
%
\bibliographystyle{splncs04}
\bibliography{aplas2021}

\begin{thebibliography}{10}
\providecommand{\url}[1]{\texttt{#1}}
\providecommand{\urlprefix}{URL }
\providecommand{\doi}[1]{https://doi.org/#1}

\bibitem{tactictoe_for_coq}
Blaauwbroek, L., Urban, J., Geuvers, H.: Tactic learning and proving for the
  coq proof assistant. In: {LPAR} 2020: 23rd International Conference on Logic
  for Programming, Artificial Intelligence and Reasoning, Alicante, Spain
  (2020)

\bibitem{sledgehammer}
Blanchette, J.C., B{\"{o}}hme, S., Paulson, L.C.: Extending sledgehammer with
  {SMT} solvers. In: Automated Deduction - {CADE-23} - 23rd International
  Conference on Automated Deduction, Wroclaw, Poland, 2011. Proceedings (2011)

\bibitem{judgement_day}
B{\"{o}}hme, S., Nipkow, T.: Sledgehammer: Judgement day. In: Automated
  Reasoning, 5th International Joint Conference, {IJCAR} 2010, Edinburgh, UK.
  (2010)

\bibitem{acl_book}
Boyer, R.S., Moore, J.S.: A computational logic handbook, Perspectives in
  computing, vol.~23. Academic Press (1979)

\bibitem{proof_plan}
Bundy, A.: The use of explicit plans to guide inductive proofs. In: Lusk, E.L.,
  Overbeek, R.A. (eds.) 9th International Conference on Automated Deduction,
  Argonne, Illinois, USA, May 23-26, 1988, Proceedings (1988)

\bibitem{alan1}
Bundy, A.: The automation of proof by mathematical induction. In: Robinson,
  J.A., Voronkov, A. (eds.) Handbook of Automated Reasoning (in 2 volumes), pp.
  845--911. Elsevier and {MIT} Press (2001)

\bibitem{rippling}
Bundy, A., Stevens, A., van Harmelen, F., Ireland, A., Smaill, A.: {R}ippling:
  {A} heuristic for guiding inductive proofs. Artif. Intell.  (1993)

\bibitem{zipperposition}
Cruanes, S.: Superposition with structural induction. In: Dixon, C., Finger, M.
  (eds.) Frontiers of Combining Systems - 11th International Symposium, FroCoS
  2017, Bras{\'{\i}}lia, Brazil, September 27-29, 2017, Proceedings. Lecture
  Notes in Computer Science, vol. 10483, pp. 172--188. Springer (2017).
  \doi{10.1007/978-3-319-66167-4\_10},
  \url{https://doi.org/10.1007/978-3-319-66167-4\_10}

\bibitem{coqhammer}
Czajka, L., Kaliszyk, C.: Hammer for {C}oq: Automation for dependent type
  theory. J. Autom. Reasoning  (2018). \doi{10.1007/s10817-018-9458-4}

\bibitem{isaplanner_phd}
Dixon, L.: A proof planning framework for {I}sabelle. Ph.D. thesis, University
  of Edinburgh, {UK} (2006), \url{http://hdl.handle.net/1842/1250}

\bibitem{isaplanner}
Dixon, L., Fleuriot, J.D.: Isa{P}lanner: {A} prototype proof planner in
  {I}sabelle. In: Automated Deduction - CADE-19, 19th International Conference
  on Automated Deduction Miami Beach, FL, USA, July 28 - August 2, 2003,
  Proceedings (2003)

\bibitem{tactictoe}
Gauthier, T., Kaliszyk, C., Urban, J.: Tactic{T}oe: Learning to reason with
  {HOL4} tactics. In: LPAR-21, 21st International Conference on Logic for
  Programming, Artificial Intelligence and Reasoning, Maun, Botswana, May 7-12,
  2017 (2017)

\bibitem{hollight}
Harrison, J.: {HOL} light: {A} tutorial introduction. In: Formal Methods in
  Computer-Aided Design, First International Conference, {FMCAD} '96, Palo
  Alto, California, USA, November 6-8, 1996, Proceedings. pp. 265--269 (1996)

\bibitem{ml4pg_acl}
Heras, J., Komendantskaya, E., Johansson, M., Maclean, E.: Proof-pattern
  recognition and lemma discovery in {ACL2}. In: Logic for Programming,
  Artificial Intelligence, and Reasoning - 19th International Conference,
  LPAR-19, Stellenbosch, South Africa, December 14-19, 2013. Proceedings (2013)

\bibitem{vampire_generalization}
Hozzová, P., Kovács, L., Schoisswohl, J., Voronkov, A.: Induction with
  generalization in superposition reasoning. EasyChair Preprint no. 2468
  (EasyChair, 2020)

\bibitem{jiang}
Jiang, Y., Papapanagiotou, P., Fleuriot, J.D.: Machine learning for inductive
  theorem proving. In: Artificial Intelligence and Symbolic Computation - 13th
  International Conference, {AISC} 2018, Suzhou, China, September 16-19, 2018
  (2018)

\bibitem{hipster}
Johansson, M., Ros{\'{e}}n, D., Smallbone, N., Claessen, K.: Hipster:
  Integrating theory exploration in a proof assistant. In: Intelligent Computer
  Mathematics {CICM} 2014 (2014)

\bibitem{lisp}
Jr., G.L.S.: An overview of common {L}isp. In: Proceedings of the 1982 {ACM}
  Symposium on {LISP} and Functional Programming, {LFP} 1980, August 15-18,
  1982, Pittsburgh, PA, {USA.} (1982)

\bibitem{holy_hammer}
Kaliszyk, C., Urban, J.: Hol(y)hammer: Online {ATP} service for {HOL} light.
  Mathematics in Computer Science  (2015)

\bibitem{acl2_industry}
Kaufmann, M., Moore, J.S.: An industrial strength theorem prover for a logic
  based on {C}ommon {L}isp. {IEEE} Trans. Software Eng.  (1997)

\bibitem{AFP}
Klein, G., Nipkow, T., Paulson, L., Thiemann, R.: The Archive of Formal Proofs
  (2004), \url{https://www.isa-afp.org/}

\bibitem{vampire}
Kov{\'{a}}cs, L., Voronkov, A.: First-order theorem proving and vampire. In:
  Computer Aided Verification - 25th International Conference, {CAV} 2013,
  Saint Petersburg, Russia, July 13-19, 2013. Proceedings (2013)

\bibitem{priority_search_tree}
Lammich, P., Nipkow, T.: Priority search trees. Archive of Formal Proofs
  (2019)

\bibitem{VerifyThis}
Lammich, P., Wimmer, S.: Verifythis 2019 -- polished isabelle solutions.
  Archive of Formal Proofs  (Oct 2019)

\bibitem{binomial}
Meis, R., Nielsen, F., Lammich, P.: Binomial heaps and skew binomial heaps.
  Archive of Formal Proofs  (2010)

\bibitem{waterfall}
Moore, J.S.: Computational logic : structure sharing and proof of program
  properties. Ph.D. thesis, University of Edinburgh, {UK} (1973)

\bibitem{acl2}
Moore, J.S.: Symbolic simulation: An {ACL2} approach. In: Formal Methods in
  Computer-Aided Design, Second International Conference, {FMCAD} '98, Palo
  Alto, California, USA, November 4-6, 1998, Proceedings (1998)

\bibitem{induction}
Moore, J.S., Wirth, C.: Automation of mathematical induction as part of the
  history of logic. CoRR  \textbf{abs/1309.6226} (2013),
  \url{http://arxiv.org/abs/1309.6226}

\bibitem{lean}
de~Moura, L.M., Kong, S., Avigad, J., van Doorn, F., von Raumer, J.: The lean
  theorem prover (system description). In: Automated Deduction - {CADE-25} -
  25th International Conference on Automated Deduction, Berlin, Germany (2015)

\bibitem{GitHub}
Nagashima, Y.: Data61/{PSL} (2017),
  \url{https://github.com/data61/PSL/releases/tag/v0.2.1-alpha}

\bibitem{lifter}
Nagashima, Y.: Li{F}t{E}r: Language to encode induction heuristics for
  {I}sabelle/{HOL}. In: Programming Languages and Systems - 17th Asian
  Symposium, {APLAS} 2019, Nusa Dua, Bali, Indonesia, December 1-4, 2019,
  Proceedings (2019)

\bibitem{smart_induction}
Nagashima, Y.: Smart induction for {I}sabelle/{HOL} (tool paper). In:
  Proceedings of the 20th Conference on Formal Methods in Computer-Aided Design
  – FMCAD 2020 (2020)

\bibitem{faster_smarter}
Nagashima, Y.: Faster smarter proof by induction in {I}sabelle/{HOL}. In: Zhou,
  Z. (ed.) Proceedings of the Thirtieth International Joint Conference on
  Artificial Intelligence, {IJCAI} 2021, Virtual Event / Montreal, Canada,
  19-27 August 2021. pp. 1981--1988. ijcai.org (2021).
  \doi{10.24963/ijcai.2021/273}, \url{https://doi.org/10.24963/ijcai.2021/273}

\bibitem{pamper}
Nagashima, Y., He, Y.: Pa{M}pe{R}: proof method recommendation system for
  {I}sabelle/{HOL}. In: Proceedings of the 33rd {ACM/IEEE} International
  Conference on Automated Software Engineering, {ASE} 2018, Montpellier, France
  (2018)

\bibitem{psl}
Nagashima, Y., Kumar, R.: A proof strategy language and proof script generation
  for {I}sabelle/{HOL}. In: de~Moura, L. (ed.) Automated Deduction - {CADE} 26
  - 26th International Conference on Automated Deduction, Gothenburg, Sweden
  (2017)

\bibitem{pgt}
Nagashima, Y., Parsert, J.: Goal-oriented conjecturing for {I}sabelle/{HOL}.
  In: Intelligent Computer Mathematics - 11th International Conference, {CICM}
  2018, Hagenberg, Austria, August 13-17, 2018, Proceedings. pp. 225--231
  (2018), \url{https://doi.org/10.1007/978-3-319-96812-4\_19}

\bibitem{boolean}
Nipkow, T.: Boolean expression checkers. Archive of Formal Proofs  (Jun 2014)

\bibitem{concrete_semantics}
Nipkow, T., Klein, G.: Concrete Semantics - With {I}sabelle/{HOL}. Springer
  (2014)

\bibitem{isabelle}
Nipkow, T., Paulson, L.C., Wenzel, M.: Isabelle/HOL - a proof assistant for
  higher-order logic, Lecture Notes in Computer Science, vol.~2283. Springer
  (2002)

\bibitem{dfs}
Nishihara, T., Minamide, Y.: Depth first search. Archive of Formal Proofs
  (2004)

\bibitem{imandra}
Passmore, G.O., Cruanes, S., Ignatovich, D., Aitken, D., Bray, M., Kagan, E.,
  Kanishev, K., Maclean, E., Mometto, N.: The imandra automated reasoning
  system (system description). In: Automated Reasoning - 10th International
  Joint Conference, {IJCAR} 2020, Paris, France, July 1-4, 2020, Proceedings,
  Part {II} (2020)

\bibitem{kd_tree}
Rau, M.: Multidimensional binary search trees. Archive of Formal Proofs  (2019)

\bibitem{vampire_induction}
Reger, G., Voronkov, A.: Induction in saturation-based proof search. In:
  Fontaine, P. (ed.) Automated Deduction - {CADE} 27 - 27th International
  Conference on Automated Deduction, Natal, Brazil, August 27-30, 2019,
  Proceedings. Lecture Notes in Computer Science, vol. 11716, pp. 477--494.
  Springer (2019). \doi{10.1007/978-3-030-29436-6\_28},
  \url{https://doi.org/10.1007/978-3-030-29436-6\_28}

\bibitem{Simpl-AFP}
Schirmer, N.: A sequential imperative programming language syntax, semantics,
  hoare logics and verification environment. Archive of Formal Proofs  (Feb
  2008)

\bibitem{LTL-AFP}
Sickert, S.: Linear temporal logic. Archive of Formal Proofs  (Mar 2016)

\bibitem{hol4}
Slind, K., Norrish, M.: A brief overview of {HOL4}. In: Theorem Proving in
  Higher Order Logics, 21st International Conference, TPHOLs 2008, Montreal,
  Canada, August 18-21, 2008. Proceedings (2008)

\end{thebibliography}

\end{document}